\documentclass{llncs}
\usepackage{booktabs}
\usepackage{xspace}
\usepackage{wrapfig}
\newcommand{\BfPara}[1]{{\noindent\bf #1.}\xspace}
\usepackage{tikz}

\usepackage{booktabs}
\usepackage[inline]{enumitem}
\usepackage{amsmath}
\usepackage{graphicx}
\usepackage{makecell}
\usepackage[colorlinks=true, allcolors=blue]{hyperref}
\usepackage{array}
\newcolumntype{L}[1]{>{\raggedright\let\newline\\\arraybackslash\hspace{0pt}}m{#1}}
\newcolumntype{C}[1]{>{\centering\let\newline\\\arraybackslash\hspace{0pt}}m{#1}}
\newcolumntype{R}[1]{>{\raggedleft\let\newline\\\arraybackslash\hspace{0pt}}m{#1}}

\usepackage{pgfplots}
\pgfplotsset{compat = newest}
\usetikzlibrary{patterns}

\usepackage{caption}
\usepackage{subcaption}
\usepackage{comment}

\newcommand{\etal}{{\em et al}.\xspace}

\usepackage{colortbl}
\usepackage{pgfplots}
\usepackage{pgfplotstable}
\pgfplotstableset{
    /color cells/min/.initial=0,
    /color cells/max/.initial=1000,
    /color cells/textcolor/.initial=,
    color cells/.code={%
        \pgfqkeys{/color cells}{#1}%
        \pgfkeysalso{%
            postproc cell content/.code={%
                \begingroup
                \pgfkeysgetvalue{/pgfplots/table/@preprocessed cell content}\value
\ifx\value\empty
\endgroup
\else
                \pgfmathfloatparsenumber{\value}%
                \pgfmathfloattofixed{\pgfmathresult}%
                \let\value=\pgfmathresult
                \pgfplotscolormapaccess
                    [\pgfkeysvalueof{/color cells/min}:\pgfkeysvalueof{/color cells/max}]%
                    {\value}%
                    {\pgfkeysvalueof{/pgfplots/colormap name}}%
                \pgfkeysgetvalue{/pgfplots/table/@cell content}\typesetvalue
                \pgfkeysgetvalue{/color cells/textcolor}\textcolorvalue
                \toks0=\expandafter{\typesetvalue}%
                \xdef\temp{%
                    \noexpand\pgfkeysalso{%
                        @cell content={%
                            \noexpand\cellcolor[rgb]{\pgfmathresult}%
                            \noexpand\definecolor{mapped color}{rgb}{\pgfmathresult}%
                            \ifx\textcolorvalue\empty
                            \else
                                \noexpand\color{\textcolorvalue}%
                            \fi
                            \the\toks0 %
                        }%
                    }%
                }%
                \endgroup
                \temp
\fi
            }%
        }%
    }
}

\usepackage{times}

\title{The Infrastructure Utilization of Free Contents Websites Reveal their Security Characteristics}
\author{Mohamed Alqadhi\inst{1}\and David Mohaisen\inst{1}}
\institute{University of Central Florida}

\begin{document}
\maketitle

\begin{abstract}

Free Content Websites (FCWs) are a significant element of the Web, and realizing their use is essential. This study analyzes FCWs worldwide by studying how they correlate with different network sizes, cloud service providers, and countries, depending on the type of content they offer. Additionally, we compare these findings with those of premium content websites (PCWs). Our analysis concluded that FCWs correlate mainly with networks of medium size, which are associated with a higher concentration of malicious websites. Moreover, we found a strong correlation between PCWs, cloud, and country hosting patterns. At the same time, some correlations were also observed concerning FCWs but with distinct patterns contrasting each other for both types. Our investigation contributes to comprehending the FCW ecosystem through correlation analysis, and the indicative results point toward controlling the potential risks caused by these sites through adequate segregation and filtering due to their concentration.

\end{abstract}


\section{Introduction} \label{sec:Introduction}
The Web has revolutionized the way users spend their time online accessing various types of content, such as books, games, music, movies, and software. For example, many game websites offer users free or paid games. Generally, websites are grouped into two groups.
\begin{enumerate*}
\item  Website content is available for a fee, and these types of websites are known as premium content websites (PCW).
\item Website content for free, where they are known as free content websites (FCWs).
\end{enumerate*} Previous studies~\cite{AlabduljabbarM22,AlqadhiATSNM22} reported that the FCWs tend to be riskier than PCWs in terms of user privacy and security features~\cite{AkhaweBLMS10,AlabduljabbarMAJCM22,AlabduljabbarMCJCM22,AlabduljabbarM22,AlqadhiATSNM22,AlrawiM16,KosbaMWTK14}, although a clear understanding of what contributes to this risk is unclear. Given the popularity of these websites and the associated risk~\cite{AlabduljabbarMAJCM22,AlabduljabbarMCJCM22,AlabduljabbarM22}, we set out to investigate the network characteristics and the hosting patterns for these websites, including the network size, the cloud service provider (CSP), and the hosting country. We do so to identify the correlation between the security features of those websites and their characteristics in terms of hosting patterns. 

\BfPara{Approach} For a complete characterization of the FCW hosting infrastructure and associated patterns, we continue to pursue the following. 
\begin{enumerate*}
\item We identify the size of networks these websites use for hosting in small, medium, and large sizes. 
\item We identify and investigate the cloud service providers for these websites and their characteristics.  
\item We study the main hosting countries of FCWs to provide a sufficient description of their hosting characteristics. 
\item We perform a correlation analysis to distinguish the security and hosting patterns for FCWs compared to PCWs.
\item We provide a correlation analysis of the hosting patterns of FCW and PCW and their security assessment.
\end{enumerate*}
In doing this correlation analysis, and in contrast to the PCWs, we hope to shed light on the features that contribute most to explaining such websites' security and privacy risks. 

Revealing the correlations between the hosting countries with hosted FCWs and PCWs determines the appropriate action governments should take to improve hosting requirements. Revealing the correlations of malicious websites with the hosting countries will focus the efforts of governments to 
\begin{enumerate*}
 \item evaluate their security standards,
 \item take a step forward in implementing more stringent security standards to combat malicious websites,
 \item and protect the end users by reviewing the privacy and security policies or mutual agreements that any website operating must adhere to in these countries.   
\end{enumerate*}

\BfPara{Contributions} We used a data set that included 1,562 FCW and PCW obtained from the research work of Alabduljabbar~\etal\cite{AlabduljabbarAMM21}. Using Pearson's correlation analysis, we examined the connections between FCW, PCW, network size, hosting CSP, and nations. We analyzed these correlations to find patterns and affinities related to various hosting arrangements. The links between website attributes, network size, hosting providers, and regional distribution are better understood due to this investigation. {\bf (1) Full Comparison.} We provide a comprehensive understanding of the different characteristics of the FCW hosting pattern compared to PCW by studying their correlations with network size, hosting CSPs, and hosting countries. {\bf (2) Systematic Analysis.} We provide a systematic security analysis for FCW and PCW. Analyze the correlations between FCW, PCW, and malicious or benign attributes of content categories.
We study the correlations of malicious FCWs and PCWs with hosting infrastructures.  {\bf (3) Hosting Correlations} We provide a detailed discussion of different characteristics of the hosting pattern. Derived from the results of a correlation analysis between small, medium, and large sizes of the networks and malicious or benign websites. We provide correlation results of the top hosting CSPs and countries. We discuss whether a strong or weak correlation exists between hosting patterns for specific content categories or security behaviors.

\BfPara{Paper Organization}
The rest of the paper reviews the related work~\ref{sec:related}, followed by research questions, data collection, and the analysis method described in Section~\ref{sec:Methodology}. The results of the analysis are given in Section~\ref{sec:Correlation}.  The detailed discussion is provided in Section~\ref{sec:discussion}. Finally, the concluding remarks and the work summary are in Section\ref{sec:final}.

\section{Related Work}\label{sec:related}
This work provided a detailed correlation analysis for FCWs and PCWs with their different networking hosting patterns and security aspects. Security analysis on FCWs has been established previously by Alabduljabbar~\etal\cite{AlabduljabbarAMM21,AlabduljabbarMAJCM22,AlabduljabbarMCJCM22,AlabduljabbarM22}. The cost of using FCWs has been investigated in~\cite{HuT17,HuTB19,LeeNHC22,RoyKN22}. The correlation between FCW security and the use of a specific content management system has been introduced in~\cite{AlqadhiATSNM22}, while other studies performed a correlation analysis on website security, such as~\cite{ChenVVPHDJ17,GoethemCNDJ14,MekovecH12,MezzourCC17,RaponiP20}. Taking into consideration the number of studies and the lack of space, we concentrate solely on a subset of relevant studies to this work and their results.

\BfPara{Security Analysis} Zhao \etal~\cite{ZhaoZL22} investigated the impacts of user-generated content (UGC) and marketer-generated content (MGC) on free content consumption by integrating the literature with research on determinants of physical exercise. 
Drutsa \etal~\cite{DrutsaGS16} investigated the utility of new data sources to predict video popularity without reliable data from video hosting services. Vasek \etal~\cite{VasekWM16} examined the effectiveness of sharing abuse data with web hosting providers to mitigate malicious online activities. 
Mirheidari \etal~\cite{MirheidariAKJ12} devised two attacks against web servers exploiting the improper isolation between files on shared web hosting servers. Also, in~\etal~\cite{MirheidariAKJ18} outlined a comprehensive overview of common attacks on shared Web servers.

\BfPara{Correlation Analysis} Several works performed a correlation analysis of the website's security. Visschers~\etal\cite{ChenVVPHDJ17} explores the cost of cybercrime and its relationship to the web security posture. Mezzour~\etal\cite{MezzourCC17} examines the relationship between social and technological factors and international variations in network-based attacks and hosting. Moreover, Mekovec ~\etal\cite{MekovecH12} found how user perceptions of security and privacy impact their evaluation of online services using correlation analysis. 

\BfPara{Domestic Analysis} Goethem~\etal\cite{GoethemCNDJ14} presents a large-scale security analysis of 22,851 websites originating in 28 European countries. Furthermore, Raponi and Di Pietro~\cite{RaponiP20} analyzed the password recovery management mechanism of Alexa's top 200 websites, with domains registered in certain European countries. They found that more than 54\% of the websites in France, 36\% in Italy, 47\% in Spain, and 33\% in the UK were vulnerable in December 2017.  
\section{Methodology}\label{sec:Methodology}

\subsection{Research Questions}
This work aims to derive insightful results of FCW correlations and different hosting patterns compared to PCW. To achieve this goal, we have worked to provide valid answers to the following questions. {\bf RQ1}. What are the main differences between the hosting patterns (networks, hosting CSPs, and countries) of FCWs compared to PCWs? {\bf RQ2}. What type of correlation exists between hosting patterns (networks, hosting CSPs, and countries) and malicious FCWs or PCWs? {\bf RQ3} What are the main correlations of the hosting patterns (networks, hosting CSPs, and countries) of content websites? {\bf RQ4}. What are the implications of FCWs hosting patterns correlation analysis?

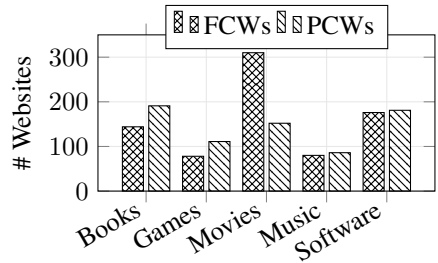
\begin{wrapfigure}{r}{0.5\textwidth} 
\pgfplotsset{width=2cm,compat=1.5}  
       \centering
    \begin{tikzpicture}  
          
        \begin{axis}  
        [  
            ybar, 
            ymin = 50, ymax = 300,
            ytick distance = 100,
            enlargelimits=0.2,  
            width = 0.9\textwidth,
            height = 0.3\textwidth,
            ylabel={\# Websites}, 
            symbolic x coords={Books, Games, Movies, Music, Software}, 
            xtick=data,  
            x=0.8cm,
            bar width=2.8mm,
            legend style={at={(0.2,1.05)},anchor=west, legend columns=2},
            xticklabel style={rotate=30,anchor=east, align=right},
            grid=both,grid style={line width=.1pt, draw=gray!10},major grid style={line width=.1pt,draw=black!10}
            ]
        \addplot[color=black,pattern=crosshatch]  coordinates {(Books,144) (Games,78) (Movies,310) (Music,80) (Software,176)};  
         \addlegendentry{FCWs};
        \addplot[color=black, pattern=north west lines] coordinates {(Books,191) (Games,111) (Movies,152) (Music,86) (Software,181)};  
        \addlegendentry{PCWs};

        \end{axis}  
        \end{tikzpicture}  \vspace{-5mm}
\caption{FCWs vs. PCWs.}
    \label{fig:dis_fcw_pcw}\vspace{-5mm}
\end{wrapfigure}

\begin{table}[t]
\centering
\caption{Network sizes and their characteristics. The maximum slash bit is 32 (IPv4). $x$ represents the number of bits and $y$ represents the number of addresses. 
}\label{tab:NetworkClassification}
\scalebox{0.8}{\begin{tabular}{lll}
\Xhline{1\arrayrulewidth}
\Xhline{1\arrayrulewidth}
{\bf Size} & {\bf Bits in CIDR} & {\bf \# Addresses}\\
\Xhline{1\arrayrulewidth}

Small (SN) & $/24<x\leq/32$ & $2^8>y\geq2^0$ \\

\Xhline{1\arrayrulewidth}
Medium (MN) & $/16< x\leq /24$ & $2^{16}>y\geq 2^{8}$ \\

\Xhline{1\arrayrulewidth}
Large (LN) & $/8<x\leq/16$ & $2^{24}>y\geq2^{16}$ \\

\Xhline{1\arrayrulewidth}
Very Large (VLN) & $/0< x\leq/8$ & $2^{32}> y\geq2^{24}$ \\

\Xhline{1\arrayrulewidth}
\Xhline{1\arrayrulewidth}
\end{tabular}}\vspace{-5mm}
\end{table}

\subsection{Data Collection Process}
We comprehend the research questions by examining multiple datasets. 
\begin{enumerate*}
    \item a main dataset of FCWs, PCWs, and their annotations, 
    \item complementary dataset for augmenting the analysis of the main dataset in terms of security (maliciousness detection),
    \item network size classification and,
    \item hosting patterns (network, CSP, and country) annotations.
\end{enumerate*}
In the following, we review these datasets.  

\BfPara{Free and Premium Websites} \label{sec:FCWs_and_PCWs_Data} We use the dataset of Alabduljabbar~\etal~\cite{AlabduljabbarMAJCM22,AlabduljabbarMCJCM22,AlabduljabbarM22}. The main criteria for selecting the sample were determined based on popularity, main language, and activities. During the collection time, all selected websites were live. Data were collected using three search engines (Bing, DuckDuckGo, and Google). The classification of content type has been applied manually, whether the website is in FCW, PCW, or (book, game, movie, music, or software) content category.

\BfPara{Malicious Annotation}\label{sec:Malicious_Classification} After collecting the data, the \href{https://www.virustotal.com/gui/home/upload}{VirusTotal} ~\cite{VirusTotal} API has been used to determine the security of each website, which is a tool that combines more than 70 scanning engines and is available online. VirusTotal enabled us to detect malicious IPs, domains, or URLs correlated with websites. We used VirusTotal since it is the standard tool used in this domain~\cite{ThomasM14,10.1007/978-3-031-26303-3_12,baek2018ssd,Mohaisen15,MohaisenAM15,SaadKM19,WangMCC15} We broadened the data collected according to the VirusTotal output. Since it gives multiple detection results, we take an entity, website, or IP as malicious if at least one of the returned scan results is at least.

\BfPara{Hosting Patterns Annotation}\label{sec:Hosting_Pattern_Annotation} We analyze the scope of the network infrastructure associated with FCWs using the IP addresses connected to each domain as a feature for analysis. We rely on two major API services--ipdata~\cite{ipdata} and IPSHU~\cite{ipshu}--to gather pertinent information about the given IP address. The subnet mask is used to determine the size of each website's network. Using the CIDR (Classless Inter-Domain Routing) notation, we classified: small networks (/25 - /32), medium networks (/16 - /24), large networks (/8 -/15), and (anything below /7) very large networks as in Table~\ref{tab:NetworkClassification}.

The IPs of FCWs and PCWs are used to determine the hosting CSPs by querying ipdata~\cite{ipdata} and IPSHU~\cite{ipshu}. They give the CSP name for the hosting site and its longitude and latitude to determine the hosting country for each website. After refining the websites, we found that only 1,509 (96. 6\%) websites are online, as appears in Figure~\ref{fig:dis_fcw_pcw}. Among the findings, 788 FCWs and 721 PCWs were grouped into five categories: books (144 free, 191 premium), games (78 free, 111 premium), movies (310 free, 152 premium), music (80 free, 86 premium), and software (176 free, 181 premium).

\subsection{Correlation Analysis}

We aim to determine whether there is a correlation between the distribution of malicious or benign FCWs and PCWs in different network sizes, CSPs, and counties. To quantify the strength of our correlation, we will use the Pearson correlation coefficient, which is calculated using the following formula: $\rho_{X,Y}=\frac{\text{cov}(X,Y)}{\sigma_X\sigma_Y}$ Here, $X$ represents free, premium, malicious, or benign attributes. On the contrary, $Y$ represents the characteristic being studied. The numerator of the formula represents the covariance between $X$ and $Y$, while the denominator represents the product of their standard deviations.

\begin{figure}
    \centering
    \includegraphics[width=0.6\linewidth]{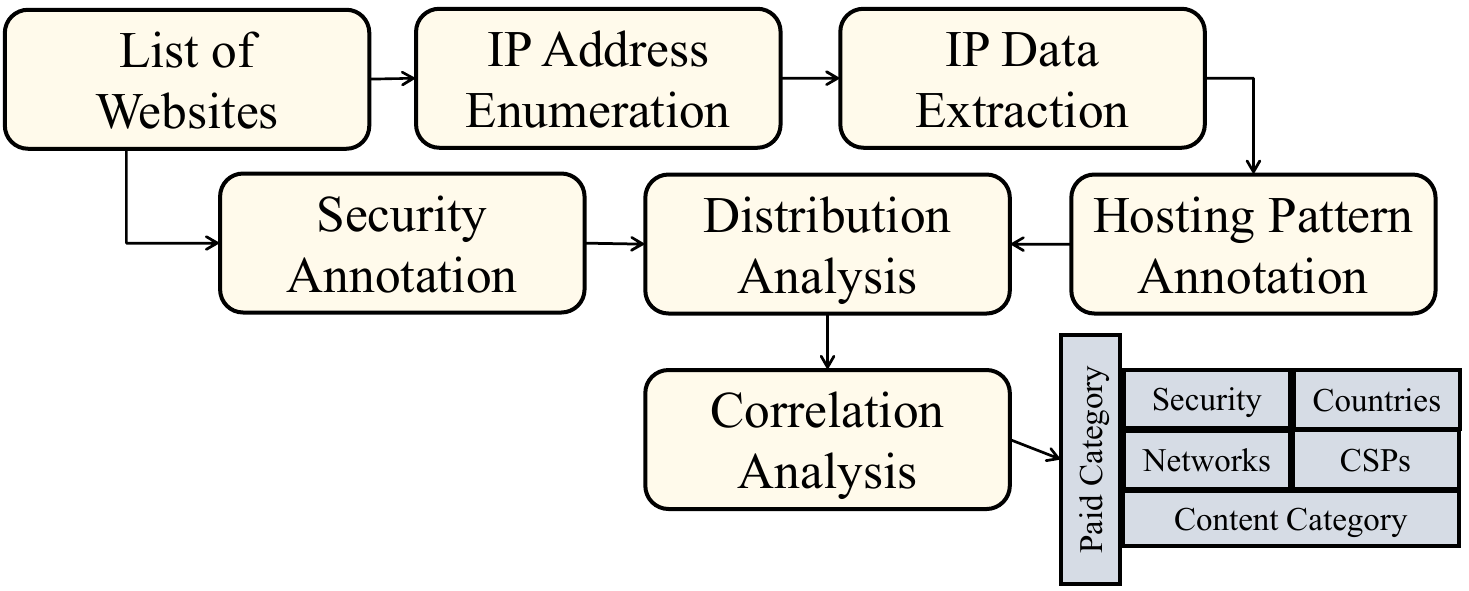}\vspace{-2mm}
    \caption{The workflow of high-level representation of our data extraction.}
    \label{fig:Analysis_Workflow}\vspace{-5mm}
\end{figure}

In this study, we used the correlation analysis approach to recognize the patterns and differences between FCW and PCW, across various analysis dimensions. This study uses six main dimensions: Type of website (FCW or PCW), type of content category, maliciousness of websites, network size, CSP, and hosting country. In the following, we define each of those dimensions as appears in the workflow of this analysis in Figure~\ref{fig:Analysis_Workflow}. 

\BfPara{Content Websites} This feature signifies the type of website, free (FCW) or paid (PCW), that correlates with a specific infrastructure entity (size, CSP, country), security attributes (malicious or benign) and content types. The paid feature of a website is determined by using different search engines, as described in the FCW and PCW data collection and annotation~\ref{sec:FCWs_and_PCWs_Data}. We study the correlation of FCWs / PCWs with different infrastructure features, security features, or content categories to determine where to focus the development effort to improve the hosting of more secure FCWs or PCWs.

\BfPara{Content Categories} In this study, the content of the websites is categorized as (books, games, movies, music, and software). We study the correlation between content types and hosting infrastructure patterns. We study their correlation with FCWs / PCWs and their security attributes (malicious or benign). To know the weaknesses of hosting different content categories. Such as the correlation of malicious websites to a specific content type in a specific hosting infrastructure.

\BfPara{Security Attributes} This feature signifies the total association of malicious or benign websites with a specific infrastructure entity (size, CSP, or country). The results of the VirusTotal scan determine the maliciousness of a website as described in the security annotation process in~\ref{sec:Malicious_Classification}. We study the correlation of malicious or benign websites with different types of content categories, the type of website (FCW, PCW), network sizes, CSPs, and countries. Inform hosting providers about the risks associated with FCW.

\BfPara{Network Size} The network size dimension represents the number of websites discovered within a specific size of the network (small, medium, large, and very large), as described in Section~\ref{sec:Hosting_Pattern_Annotation}. We provide the results of the correlation analysis between FCWs and PCWs, the content categories of websites, and malicious associations with the size of the network.  Investigating FCWs' networks is essential to know their weaknesses.

\BfPara{Cloud Service Provider} The CSP indicates the cloud service provider used to host FCW or PCW. We are studying the correlation between the top ten and the other 298 CSPs discovered during this study. We study the correlations of CSPs with the different content categories and security attributes. Determine the security policies of the CSPs that need to be investigated, altered, or improved.

\BfPara{Country} This feature represents the hosting countries obtained from the hosting patterns annotations~\ref{sec:Hosting_Pattern_Annotation}. We found 44 countries with a heavy-tailed distribution. We study the correlation between FCWs/PCWs and their malicious association for different content categories with the hosting countries. To know where we can make any improvements to the security agreements or rules of hosting FCWs.

\section{Correlation Results} \label{sec:Correlation}
We investigate the correlations between the analysis dimensions described in Section~\autoref{sec:Methodology}. Specifically, we examine the correlation between FCW and PCW with other characteristics, including network sizes, CSPs, and countries. We also study the malicious and benign classifications in their hosting infrastructures. Figures~\ref{General Networks},~\ref{fig:(Free vs Premium) CSPs Correlation}, and ~\ref{fig:Top 10 Countries Correlation}, illustrate the correlation between FCW and PCW with various features of the infrastructure that indicate the five categories of content.

\subsection{General Correlation Results}
FCWs strongly correlate with malicious websites, and most FCWs reside in medium and small networks.  PCWs are more prevalent in large networks. We also found a strong relationship between the top hosting CSPs and malicious FCWs. However, the relationship between the top countries and PCWs is more diverse. Interestingly, countries correlated with hosting FCWs are found to be more related to hosting malicious websites. The following are the most noticeable insights from the correlation analysis.

   \BfPara{Malicious or Benign} We notice that FCWs are mostly correlated with the malicious attribute. As appears in Figure~\ref{General Networks}, a negative correlation coefficient varies between 0.13 and 0.46. The strongest correlation is found on software websites. The weakest correlation is found on movie websites. Unlike the malicious website, we found a strong correlation between PCWs and benign attributes. The highest positive correlation coefficient is 0.46 on software websites, and the lowest was discovered on movie websites.

    \BfPara{Network Correlation} Per Figures~\ref{General Networks} and \ref{fig:Malicious VS Benign Network}, we observe a strong correlation between FCW and medium networks, especially in games, movies, and music websites. There is a weak correlation between small networks and FCWs. Compared to the large network that strongly correlates with PCWs. Noticeably, there are no correlations with very large networks. Moreover, benign attributes are correlated with large and small networks. The malicious attribute has a strong correlation with medium networks.
    
    \BfPara{CSPs Correlations} The top ten CSPs strongly correlate with the FCWs. Although some CSPs show a strong correlation with premium websites, most CSPs correlate strongly with benign attributes. Only two of the top ten CSPs have a strong correlation with malicious websites. Some of the content categories in the top ten CSPs indicate a weak correlation with malicious websites, as we can see in Figures~\ref{fig:(Free vs Premium) CSPs Correlation}, and \ref{fig:(Malicious vs Benign) CSPs Correlation}.

    \BfPara{Countries Correlations} The top hosting countries strongly correlate with benign websites. Similar to the other countries. In the opposite direction, some content categories indicate a strong correlation to malicious websites in several countries. Such as the games websites in the United States and Belgium as in Figure~\ref{fig:(Malicious vs Benign) Countries Correlation}. However, the FCWs strongly correlate with the United States, Germany, Australia, France, and other countries. Especially for the categories of books, movies, and software as in Figure~\ref{fig:Top 10 Countries Correlation}.

Furthermore, since we provided a summary of the most important findings of the results from the correlation analysis, the following will be a detailed analysis of the network size, CSPs, and countries' correlation to FCWs and PCWs.

\begin{figure}[!htb]
    \centering
    \begin{minipage}{.5\textwidth}
    \centering
    \vrule
    \pgfplotstabletypeset[
        color cells={min=-1,max=1,textcolor=black},
        /pgfplots/colormap={orangewhiteorange}{rgb255=(255,50,0) color=(white) rgb255=(0,100,255)},
        /pgf/number format/fixed,
        /pgf/number format/precision=3,
        col sep=comma,
        columns/Corr/.style={reset styles,string type}
    ]{
        Corr, (+), (-), SN, MN, LN, VLN
        Books, 0.02, -0.02, -0.08, -0.07, 0.11, 0.05
        Games, 0.32, -0.32, 0.02, -0.20, 0.21, 0.00
        Movies, 0.13, -0.13, -0.01, -0.22, 0.23, 0.00
        Music, 0.24, -0.24, -0.01, -0.18, 0.18, 0.00
        Software, 0.46, -0.46, -0.13, -0.08, 0.16, 0.00
        Overall, 0.20, -0.20, -0.05, -0.14, 0.17, 0.03
    }
    \vrule
    \captionsetup{justification=justified, width=.8\textwidth}
    \caption{General networks. Red and blue indicate a vital contribution FCWs and PCWs respectively.} 
    \label{General Networks}
    \end{minipage}~
    \begin{minipage}{0.5\textwidth}
    \centering
        \centering
        \vrule
        \pgfplotstabletypeset[
            color cells={min=-1,max=1,textcolor=black},
            /pgfplots/colormap={orangewhiteorange}{rgb255=(0,255,0) color=(white) rgb255=(255,0,0)},
            /pgf/number format/fixed,
            /pgf/number format/precision=3,
            col sep=comma,
            columns/Corr/.style={reset styles,string type}
        ]{
        Corr, SN, MN, LN, VLN
        Books, -0.04, 0.10, -0.09, -0.04
        Games, -0.12, 0.22, -0.19, 0.00
        Movies, -0.03, 0.08, -0.08, 0.00
        Music, -0.07, 0.17, -0.16, 0.00
        Software, -0.06, 0.21, -0.20, 0.00
        Overall, -0.05, 0.14, -0.13, -0.02
        }
        \vrule
        \captionsetup{justification=justified, width=.8\textwidth}
        \caption{The correlation of malicious vs. benign networks. Red for malicious and green for benign.} \label{fig:Malicious VS Benign Network} 
    \end{minipage}
\end{figure} \vspace{-10mm}

\subsection{Networks Correlations}
\BfPara{General Networks} Figure~\ref{General Networks} shows the relationship between FCWs and PCWs in different network sizes, indicating their correlation with various content categories. The results indicate a strong relationship between the book FCWs, predominantly hosted in small and medium networks. In contrast, large networks strongly correlate with PCWs. Unlike the book category, the games, movies and music categories show a weak correlation with small networks, which varies between FCWs and PCWs.

 Figure~\ref{General Networks} illustrates the associations between FCWs and PCWs with different networks. The correlation highlights the connections between FCWs and PCWs in different categories that use different network sizes and their malicious or benign classification. Reflecting the network correlation observed earlier in Figure~\ref{General Networks}, malicious attributes emerge as a significant factor. However, the distinction here is the pronounced association between malicious attributes and FCWs, suggesting that medium and small networks are more strongly linked to malicious factors than other websites.

\BfPara{Malicious Networks} The heat map shown in Figure~\ref{fig:Malicious VS Benign Network} illustrates the associations between malicious websites and various characteristics of the size of the network. This correlation highlights the connections between malicious and benign websites in different categories and network sizes. For example, red indicates a high concentration of malicious websites and green represents predominantly benign content. On examination, we observe that most malicious websites are found in medium-sized networks, while small and large networks mainly consist of benign websites. Furthermore, there is a notable correlation between the malicious attribute and the category of games in medium networks, with similar patterns observed on music and software websites. On the contrary, the ``Other'' categories exhibit a weaker likelihood of maliciousness.

\subsection{Cloud Service Providers}
\BfPara{General Correlations} Figure~\ref{fig:(Free vs Premium) CSPs Correlation} illustrates the associations between FCWs and PCWs and the CSPs most commonly used in the top hosting countries. The correlation shows the connections between FCWs and PCWs over the top-used CSPs. Furthermore, we found that most PCWs are associated with CSPs that report the lowest malicious activity. We notice that FCWs are used primarily with the most malicious websites that host CSPs, as appears in Figure~\ref{fig:(Malicious vs Benign) CSPs Correlation}. In contrast, PCWs are used primarily with the least malicious websites hosting CSPs. Most PCWs, whose categories are games, movies, and music, are hosted by ``Amazon", while ``Cloudflare" hosts most FCWs of books and software FCWs. Finally, the general categories are highly distributed.

\begin{figure}[t]
\centering
\vrule
\pgfplotstabletypeset[
    color cells={min=-1,max=1,textcolor=black},
    /pgfplots/colormap={orangewhiteorange}{rgb255=(255,50,0) color=(white) rgb255=(0,100,255)},
    /pgf/number format/fixed,
    /pgf/number format/precision=3,
    col sep=comma,
    columns/Corr/.style={reset styles,string type}
]{

Corr, CF, AZ, LW, TR, GO, ST, LS, AK, FS, MS, Or
Books, -0.07, 0.19, -0.16, -0.16, 0.14, -0.14, -0.06, 0.1, 0.11, 0.07, -0.04
Games, -0.2, 0.26, -0.15, -0.09, 0.12, 0.00, -0.15, 0.21, 0.14, 0.06, -0.07
Movies, -0.17, 0.39, -0.2, -0.18, 0.09, -0.16, -0.15, 0.18, 0.15, 0.08, 0.06
Music, -0.17, 0.39, -0.05, -0.11, -0.01, -0.2, -0.08, 0.11, 0.06, 0.04, -0.1
Software, -0.27, 0.13, -0.18, -0.09, 0.07, 0.00, -0.17, 0.13, 0.07, 0.16, 0.2
Overall, -0.15, 0.26, -0.18, -0.16, 0.08, -0.15, -0.15, 0.15, 0.11, 0.09, 0.05

}
\vrule
\caption{Most used CSP's analysis. The color indication is similar to~\ref{General Networks}. The top hosting CSPs are, ``Cloudflare"(CF), ``Amazon"(AZ), ``Liquid Web"(LW), ``Trellian"(TR), ``Google"(GO), ``Sp-Team"(ST), ``LeaseWeb"(LS), ``Akamai"(AK), ``Fastly"(FS), ``Microsoft"(MS), Other CSPs(Or).}
\label{fig:(Free vs Premium) CSPs Correlation}
\vspace{-4mm}
\end{figure}

\BfPara{Malicious Correlations} Figure~\ref{fig:(Malicious vs Benign) CSPs Correlation} shows the relationship between malicious websites and the top hosting CSPs. The correlation indicates the relation between malicious and benign websites on the most used CSPs. We notice that the highest concentration of malicious websites strongly correlates with the CSPs ``Cloudflare" and ``Liquid Web", while benign websites are primarily associated with the other CSPs. Interestingly, book websites exhibit a strong malicious relationship with ``Microsoft" CSP, which is known to have one of the lowest reported percentages of malicious activity. The Movies also display multiple malicious correlations with the top six CSPs compared to the others.

\begin{figure}[t]
\centering
\vrule
\pgfplotstabletypeset[
    color cells={min=-1,max=1,textcolor=black},
    /pgfplots/colormap={orangewhiteorange}{rgb255=(0,255,0) color=(white) rgb255=(255,0,0)},
    /pgf/number format/fixed,
    /pgf/number format/precision=3,
    col sep=comma,
    columns/Corr/.style={reset styles,string type}
]{
Corr, CF, AZ, LW, TR, GO, ST, LS, AK, FS, MS, Or
Books, 0.58, -0.2, -0.02, -0.09, -0.06, -0.08, -0.03, -0.01, -0.08, 0.12, -0.28
Games, 0.7, -0.22, 0.14, 0.08, -0.13, 0.00, 0.06, -0.22, -0.15, -0.07, -0.44
Movies, 0.11, -0.15, 0.05, 0.02, 0.04, 0.08, 0.02, -0.05, -0.06, -0.02, -0.05
Music, 0.52, -0.17, 0.12, -0.07, -0.08, -0.05, -0.05, -0.07, -0.12, -0.08, -0.16
Software, 0.58, -0.16, 0.15, -0.01, -0.09, 0.00, -0.04, -0.11, -0.1, -0.13, -0.38
Overall, 0.48, -0.18, 0.06, -0.03, -0.05, -0.01, -0.01, -0.08, -0.09, -0.04, -0.25
}
\vrule
\caption{Malicious vs. benign hosting CSPs. The colors are similar to~\ref{fig:Malicious VS Benign Network}.}
\label{fig:(Malicious vs Benign) CSPs Correlation}
\vspace{-6mm}
\end{figure}

\subsection{Countries Correlation}
\BfPara{General Correlations} Figure~\ref{fig:Top 10 Countries Correlation} shows the relationship between FCW and PCW in the top 10 hosting countries. The correlation indicates the relationship between FCW and PCW in the top hosting countries. For example, we noticed a strong relationship between PCWs in the movie category and the United States and fewer correlations with other categories. Moreover, we observed that most of the top hosting countries exhibit strong relationships with FCWs, especially those reported to be highly malicious. For example, countries such as China, the UK, Canada, and Ireland show weak relationships with PCWs, but surprisingly, it is more vital than their relationship with the FCWs. Furthermore, we noticed a high concentration of FCWs in the game and software categories in Belgium, which are reported to be the most malicious websites. Simultaneously, a strong association can be observed between FCWs in movies and music categories and Germany, which is reported to have a low level of malicious activity.

\begin{figure}[t]
\centering
\vrule
\pgfplotstabletypeset[
     color cells={min=-1,max=1,textcolor=black},
     /pgfplots/colormap={orangewhiteorange}{rgb255=(255,50,0) color=(white) rgb255=(0,100,255)},
    /pgf/number format/fixed,
    /pgf/number format/precision=3,
    col sep=comma,
    columns/Corr/.style={reset styles,string type}
]{
Corr, US, BE, NL, DE, AU, FR, CN, GB, CA, IE, Or
Books, 0.03, -0.17, 0.04, -0.17, -0.13, -0.06, 0.13, 0.11, 0.11, 0.08, -0.03
Games, 0.1, -0.25, 0.11, 0.00, -0.02, 0.03, 0.03, 0.09, 0.09, 0.09, -0.2
Movies, 0.29, -0.16, -0.07, -0.22, -0.15, -0.07, 0.15, 0.09, 0.00, 0.16, -0.08
Music, 0.14, -0.1, -0.04, -0.25, -0.05, 0.04, 0.11, 0.13, -0.08, 0.15, -0.07
Software, 0.19, -0.29, -0.08, 0.04, -0.05, 0.01, 0.07, -0.05, 0.04, 0.09, 0.03
Overall, 0.17, -0.19, -0.03, -0.16, -0.13, -0.02, 0.11, 0.07, 0.05, 0.12, -0.04
}
\vrule
\caption{Top hosting countries (Alpha-2). The colors are similar to~\ref{General Networks}.}
\label{fig:Top 10 Countries Correlation}
\vspace{-2mm}
\end{figure}

\BfPara{Malicious Correlations} Figure~\ref{fig:(Malicious vs Benign) Countries Correlation} shows the correlation between malicious and benign websites with the top ten hosting countries. The correlation highlights the relationship between malicious and benign websites in the top hosting countries. The heat map reveals a strong relationship between the United States and Belgium that hosts malicious websites, particularly on books, games, and software websites. Most other countries have a strong connection to benign websites. 
\begin{figure}[t]
\centering
\vrule
\pgfplotstabletypeset[
    color cells={min=-1,max=1,textcolor=black},
    /pgfplots/colormap={orangewhiteorange}{rgb255=(0,255,0) color=(white) rgb255=(255,0,0)},
    /pgf/number format/fixed,
    /pgf/number format/precision=3,
    col sep=comma,
    columns/Corr/.style={reset styles,string type}
]{
Corr, US, BE, NL, DE, AU, FR, CN, GB, CA, IE, Or
Books, 0.11, 0.12, -0.09, -0.13, -0.09, 0.06, -0.01, 0.02, 0.02, -0.05, -0.08
Games, 0.31, 0.27, -0.17, -0.15, 0.01, -0.04, -0.1, -0.16, -0.09, -0.09, -0.21
Movies, -0.06, 0.05, 0.01, 0.02, 0.01, 0.04, -0.03, -0.04, -0.04, -0.06, 0.08
Music, 0.11, 0.2, -0.09, 0.03, -0.08, 0.02, 0.05, -0.08, -0.1, -0.1, -0.11
Software, -0.04, 0.29, -0.08, -0.12, -0.03, 0.05, -0.06, 0.05, -0.11, -0.08, -0.1
Overall, 0.05, 0.2, -0.07, -0.07, -0.04, 0.04, -0.03, -0.01, -0.05, -0.07, -0.06
}
\vrule
\caption{Malicious vs. benign hosting countries (Alpha-2). Colors are similar to~\ref{fig:Malicious VS Benign Network}.}
\label{fig:(Malicious vs Benign) Countries Correlation}
\vspace{-6mm}
\end{figure}

\section{Results Discussion}\label{sec:discussion}
In this section, we will discuss the key insights of the correlation analysis. Highlighting the answers to the research questions. We will list the challenges we encountered during the study. Finally, we will shed light on the limitations and recommendations. 
\subsection{Results Takeaway}
To sum up the key results of the correlation analysis, we will highlight the insights that provide detailed answers to the research questions as follows.

\BfPara{Free or Premium} The correlation analysis results answer {\bf RQ1}.

\begin{enumerate*}
    \item  We notice the difference in the hosting patterns of FCWs and PCWs in their common network size, the top CSPs, and most of the hosting countries.
    \item FCWs have a weak correlation to small networks, whereas PCWs have no correlation to small networks.
    \item  FCWs have a strong correlation with ``Cloudflare", ``Liquid Web", ``Trilian", ``SP-Team", and ``LeaseWeb" CSPs. Although PCWs seem to have a strong correlation with the other top ten hosting CSPs.
    \item Some of the top hosting countries show a strong correlation with FCWs. On the contrary, the other top hosting countries have a strong correlation with PCWs.
    \item The results of FCWs depict certain hosting patterns that are uniquely different from PCWs. Indicating the differences in their security behavior.
\end{enumerate*}

\BfPara{Malicious or Benign} The results also provide answers to {\bf RQ2} where we find the network hosting patterns for malicious websites.
\begin{enumerate*}
    \item Malicious websites show a strong association with FCW, while benign websites strongly correlate with PCW.
    \item In general, malicious websites have a strong correlation with the medium size of the networks. The benign websites are strongly correlated with large networks and weakly with small networks.
    \item The top ten CSPs and the other hosting CSPs show a significant correlation to benign websites. Some of the top ten CSPs have a strong correlation with hosting malicious websites.
    \item Hosting countries seem to have a significant correlation with benign websites. In the opposite direction, some of the content categories exhibit a strong correlation with malicious attributes. Especially in the top two hosting countries.
\end{enumerate*}

\BfPara{Hosting Patterns} The results of studying the different categories of website content in the different hosting patterns give significant answers to {\bf RQ3}.
\begin{enumerate*}
    \item Different content categories show a different level of correlation to malicious and benign attributes. Games and software websites exhibit a strong correlation with malicious FCWs. 
    \item Small networks have a weak correlation with all FCW content categories. Medium networks have a strong correlation with FCW games, movies, and music websites. Large networks show a lower correlation with such categories in PCWs.
    \item We notice the differences in the correlation with the top hosting CSPs. We found a strong correlation between all content categories and FCWs in the top hosting CSPs. Other top CSPs strongly correlate with PCWs. Such as ``Amazon", ``Akamai", ``Fastly", and ``Microsoft".
    \item There is a weak correlation between all categories of PCWs and some of the top hosting countries. In contrast, we found a strong correlation of FCWs and categories hosted in ``Belgium". 
\end{enumerate*}

\BfPara{Results Implications} The implications of the previous findings provide answers to {\bf RQ4}.
\begin{enumerate*}
    \item Isolation of FCWs that use small networks may be considered an applicable solution to mitigate FCW and PCW risks. 
    \item  Addressing malicious environments within CSPs is one of the most effective solutions to reduce risk exposure. 
    \item Taking legal action to force such CSPs to improve their security could be a viable solution to secure the network. 
     \item FCWs and PCWs are concentrated in medium networks, the same as malicious content websites. This implies the need for a better solution than isolating these networks.
     \item Games and software content are the most correlated with malicious websites. This implies the serious need to develop security scanning tools specialized in detecting malicious code that may be injected into software or game FCWs.
\end{enumerate*}

\subsection{Limitations and Recommendations}
\BfPara{Limitations} Initially, our main data set consisted of 1,562 FCW and PCW. However, after the network annotation process, we found that only 1,509 websites were operating, suggesting a decrease over time.  Thus, longitudinal analysis is required to gain insight into changes in an operation performed on these websites. The top hosting CSPs discovered during this study are widely spread. Where some of these CSPs have different companies, we combined all of the companies of the same entity into one CSP. For example, ``Amazon" CSPs provide their services regionally, such as Amazon Data Services Canada and Amazon Data Services France. Consequently, all these CSPs were aggregated into one entity ``Amazon". For further analysis of their service distribution, it is imperative to conduct further investigation to ensure their security.

\BfPara{Recommendations} Based on the findings, our recommendations to system administrators are to apply stronger security protocols. To protect their networks from malicious activities. In particular, organizations must prioritize segmenting medium-sized networks as they are often malicious. Moreover, analyzing the CSPs used by FCWs and PCWs can aid in determining which CSPs have a higher number of malicious websites than good ones. This indicates where legal action need to be considered if necessary. General observation suggests that improvements should be made by developing these aspects, reducing malicious websites, and strengthening overall network security.

\section{Conclusion} \label{sec:final}
The correlations between FCWs, PCWs and their hosting habits in network size, CSP, and hosting countries have been revealed by this research. Our investigation has shown a significant association between FCWs and medium networks, suggesting that these networks tend to host malicious websites. Additionally, we have identified some CSPs that may need to increase their security requirements, as they are significantly correlated with hosting more malicious content categories. Furthermore, our research shows a notable association between the nations where FCWs are hosted, pointing to the need for more stringent laws and other countermeasures to address dangerous websites.

\end{document}